\documentstyle[12pt]{article}
\def\d{\delta}

\def\v{\varepsilon}
\begin{document}
\vspace{20mm}

\title{Anomalies of Transport \\ in Reflectionally Noninvariant Turbulence}
\author{\sl A.V.Chechkin, V.V.Yanovsky,\\
Laboratory for Turbulence Research \\
Institute for Single Crystals National Acad.Sci.Ukraine,\\
Lenin ave. 60, Kharkov 310001, Ukraine\\
and\\
\sl A.V.Tur\\
Observatoire Midi--Pyrenees, 14, av.
Edouard--Belin,\\ F31400 Toulouse, France}
\date{}
\maketitle

\begin{abstract}
We consider the transport of passive admixture in locally homogeneous
isotropic reflectionally noninvariant turbulence of incompressible fluid.
It is  shown that anomalous convective flow appears which direction
does not coincide with that of a mean flow.
\end{abstract}

\section{Introduction}

There are many papers devoted to the study of passive  admixture transport
in a turbulent medium (see, e.g., [1, 2, 3]). The interest to this problem
is explained by its importance and deep theoretical tasks connected with
its solution. To solve the problem it is necessary to advance in
understanding the problems connected with the turbulence.

The theory of the transport in a homogeneous and isotropic turbulence has
been developed best of all. This is explained by the success in
understanding this type of the  turbulence. One of the first results,
which is already classical now, is Richardson law [4] (4/3 power), which
determines the relative diffusion coefficient in Kolmogorov turbulence
[5]. The appearance of new ideas, connected with fractal nature of
dissipation zones [6--8] leads to the generalization of the Richardson law
[9,10].

Recently, the methods of the field theory were intensively applied to the
calculation of anomalous transport coefficients in homogeneous isotropic
turbulence. The renormalization group method seems to be most effective
[11, 12].

In general, turbulent transport is determined by two principally different
effects. First of all, this is a diffusion extension of an admixture
"cloud". Such an extension is due to the  appearance of the diffusion flow
$\vec{J_D} = -D_T \nabla \langle C\rangle $, where $\langle C\rangle $ is
the mean concentration of the passive admixture. The coefficient $D_T $ of
concentration gradient (in general case it is a tensor) is the turbulent
diffusion coefficient. It determines quantitative characteristics of the
extension of the passive admixture cloud.

The transport of the cloud as a whole with changing its center position is
determined by the convective flow $\vec{J_e} = \vec{V} \langle C\rangle
$, where $\vec{V}$ is the transport velocity which determines the value
and the direction of the convective flow of passive admixture.

In a homogeneous isotropic turbulence there is no convective flow, and the
evolution of a passive scalar is determined by the diffusion flow
only.  A lot of papers is devoted to the calculation of turbulent
diffusion coefficients just for this case (see, e.g., [2, 3, 13--15]. The
case of a locally homogeneous isotropic turbulence with $\langle
\vec{V}\rangle = const $, actually, is reduced to the previous case when
one passes to a reference frame moving with the velocity $\langle
\vec{V}\rangle $.  In this case the convective flow is trivial, $
\vec{J_e}=\langle \vec{V}\rangle \langle C\rangle $, and the transport of
the cloud as whole is determined by the value and the direction of
$\langle \vec{V}\rangle $.

More interesting and nontrivial effects in the convective transport should
be expected, at least, in the case of a weak spatial dependence of
$\langle \vec{V}\rangle $. In this case there is no possibility to pass to
single reference frame moving with the velocity of the medium.

In our paper the main attention is devoted to nontrivial effects which
determine the convective flow of a passive admixture "cloud".
Nevertheless, diffusion flows are also studied.

 As it is shown in our paper, a weak dependence of $\langle \vec{V}\rangle
 $ on coordinates is not sufficient for changing the convective flow. It
 is necessary to take into account more delicate properties  of the
 turbulence. The property of helicity plays the principal role in the
 appearance of nontrivial effects. The helical turbulence is one more type
 of the turbulence. This type is characterized by the pseudoscalar
 $\langle \vec{V} curl \vec{V}\rangle $ (helicity) which differs from
 zero. The reflected symmetry is violated in the helical turbulence, this
 property being not restored in the developed turbulence [16]. The reason
 for a substantial role of the helicity is  due to its topological nature
 (see, e.g., [17--19]). The non--zero value of the helicity implies the
 linking of vortex lines of the velocity field, that is, a nontrivial
 topology of this field.

 In the presence of helical turbulence a substantial reconstruction of
 classical instabilities, e.g., convective instability [20], takes place.
 Probably, it is convective instability which the birth of typhoon is
 connected with [21].

 In this paper the peculiarities of the passive admixture transport
 arising in reflectionally noninvariant turbulence are discussed. The main
 attention is devoted to the derivation of the equation for the average
 concentration and to the qualitative effects due to helicity. It can be
 understood from the symmetric considerations, that in order for new
 effects to be realized the additional factors besides helicity should
 exist, for example, the inhomogeneous mean velocity of the medium. In
 some sense, this fact implies that we approach closer to the  problems of
 a more realistic turbulence.

Therefore, in this paper the transport in a turbulent medium with the mean
velocity  depending on coordinates is considered. Turbulent fluctuations
are assumed reflectionally  noninvariant and small--scale (comparatively
to the scale of the mean flow).

With the use of multiscale formalism the equation is obtained which
describes the  evolution of the average concentration of the passive
admixture in this turbulent medium. In  this equation there are two
anomalous terms describing turbulent diffusion and turbulent
convective transport, respectively. The turbulent diffusion coefficient
and the anomalous convective velocity depend on the mean velocity. The
coefficient of turbulent diffusion does not depend on helicity. The
isotropic reflectionally invariant turbulence leads only to the enhancing
the diffusion flows. The role of the reflection noninvariance of the
turbulence is more radical, because new convective flows of admixture are
formed in helical turbulence. It is worthewhile note that the direction of
new convective flows does not coincide with the  direction of the mean
velocity. In other words, the influence of nontrivial topology of a
turbulent velocity field manifests itself in changing the direction and
the value of the  convective transport of a passive admixture cloud. This
fact is discussed in the paper in details.

The existence of such flows can be principal for understandings the
admixture transport in astrophysical and geophysical phenomena.\\

\section{Average~~ equation~~ for the~ passive~ admixture transport}

Following conventional statement of problems concerning passive admixture
transport in a given turbulent field (so-called kinematic statement [18]),
we take the transport equation for a passive admixture as initial one
in our approach:
\begin{equation}\label{1}
{\partial C \over \partial t} + \vec{V}\nabla C = D_M \nabla C ,
\end{equation}
where $C(\vec{x},t)~$ is the density field of the passive admixture,
$D_M~$ is the molecular diffusivity, $\vec{V}(\vec{x}, t)$~ is the
incompressible velocity field, $ div \vec{V} =0$, which can be represented
in the form
\begin{equation}\label{2}
\vec{V} = \langle \vec{V}\rangle + \delta \vec{V} ~~~,
\end{equation}
where $ \langle \vec{V}\rangle  $ is the velocity of the regular flow
which is assumed to be a function of $\vec{x}, \delta \vec{V}(\vec{x},
t) $ is the turbulent velocity field, symbol $\langle ~~\rangle $
implies averaging over realizations of the stochastic velocity field.
Equation (1) can be written in a more fundamental form as a continuity
equation :
\begin{equation}\label{3}
{\partial C \over \partial t} + div \vec{J} =0 ~~~,
\end{equation}
where
\begin{equation}\label{4}
\vec{J} = \vec{V} C- D_M\nabla C
\end{equation}
is the vector of the admixture flow.

The form of transport equation (3) and (4) is the most general and
physically motivated. This form allows us to differentiate the terms
according to their physical meaning  in a correct way [22]. The first term
in Eq.(4) describes the convective transport of an admixture, that is, an
initial distribution of the admixture $ C(\vec{x}, t)$ is carried with the
speed $\vec{V}(\vec{x}, t) $, while the second term in Eq.(4) describes
the extension of the initial distribution due to molecular diffusion.

Our task is to derive the equation for the  average concentration
$\langle C \rangle $. When doing it we use natural physical assumption
that characteristic space and time scales of turbulent fluctuations are
much smaller then the characteristic space and time scales of the average
quantities. This assumption allows us to use the asymptotic method of
multiscale  expansion [23] in order to derive the average diffusion
equation. We introduce "slow" variables $\vec{X},T_i, i=1,2,...$ which
characterize average quantities, together with "fast" variables
$\vec{x},t$, which characterize small--scale turbulent fluctuations.
$\vec{x},t ~$ are connected with $\vec{X}, T_i~~$ as
\begin{equation}\label{5}
\vec{X} = \varepsilon \vec{x};~~~T_i =\varepsilon^i t ~~,
\end{equation}
and
$$ {\partial \over \partial \vec{x}} \to {  \partial  \over  \partial
\vec{x}}+ \v { \partial  \over  \partial \vec{X}}$$
$${ \partial \over \partial t} \to  {\partial \over  \partial t}+\v {
 \partial  \over  \partial T_1} +\v^2{ \partial \over  \partial T_2}+ ...
 ~~~.$$
$\v~$ is a small parameter characterizing the ratio of scales of turbulent
fluctuations and the average flow. The introduction of the single space
scale and the set of time scales implies that we restrict our
consideration by the processes on the space scale which are of the order
of that of the mean flow, nevertheless, we want to follow the evolution of
these processes for a  "sufficiently long" time.

In multiscale formalism the solution of Eq.(1) (or Eq.(3) ) can be found
in the form of a function depending on fast and slow variables, which are
assumed as independent. The solution is found in the form of asymptotic
expansion in powers of $\v~$ :
\begin{equation}\label{6}
C=\langle C \rangle + \v C_1 +\v^2 C_2 + ... ~~~,
\end{equation}
where $ \langle C \rangle $ is the average concentration which depends
only on slow variables $\vec {X}, T_i~; C_i , i=1,2,...$ depend on both
fast and slow variables. The next two equations in the hierarchy are
\begin{equation}\label{7}
O(\v ) : {\partial C_1 \over \partial t}+{\partial \langle C \rangle \over
\partial T_1} +\langle \vec{V} \rangle {\partial C_1\over \partial
\vec{x}} +\delta \vec{V} {\partial  \langle
C\rangle \over \partial \vec{X}}+\langle \vec{V} \rangle {\partial
\langle C \rangle  \over \partial \vec{X}} +\delta\vec{V}{\partial
C_1\over \partial \vec{x}} = D_M{\partial^2 C_1 \over \partial \vec{x}^2}
,
\end{equation}
$$O(\v^2) : {\partial C_2 \over \partial t}+{\partial C_1 \over \partial
T_1} +{\partial \langle C \rangle \over \partial T_2}+\langle \vec{V}
\rangle {\partial C_2 \over \partial \vec{x}}+\delta \vec{V}{\partial c_2
\over \partial \vec{x}} +\langle \vec{V} \rangle{\partial C_1 \over
\partial \vec{X}} =$$
\begin{equation}\label{8}
=2D_M {\partial ^2 C_1 \over \partial\vec{x}\partial\vec{X} } +D_M
{\partial ^2 \langle \vec{V} \rangle \over \partial \vec{X}^2} + D_M
{\partial ^2C_2 \over \partial \vec{x}^2}~~.
\end{equation}

For the mean concentration at zeroth  order  approximation the
evolution equations at slow times $T_1, T_2$ are obtained from Eqs. (7),
(8) as the solvability conditions (secular equation) :
\begin{equation}\label{9}
{\partial \langle C \rangle \over \partial T_1} + \langle \vec{V}\rangle
{\partial \langle C\rangle \over \partial \vec{X}} =0 ~~.
\end{equation}
\begin{equation}\label{10}
{\partial \langle C\rangle\over \partial T_2} = D_M {\partial^2\langle
C\rangle \over \partial \vec{X}^2} - \langle \vec{V}{\partial C_1\over
\partial \vec{X}} \rangle~~.
\end{equation}
The equation for $C_1$ obtained by subtracting Eq.(9) from Eq.(7). Then,
we solve the equation for $C_1$  formally with using the Green's function
and insert the  solution into Eq.(10). Thus, we get
\begin{equation}\label{11}
{\partial \langle  C \rangle \over \partial T_2} = D_M {\partial ^2
\langle  C \rangle \over \partial \vec{X}^2} + {\partial \over \partial
X_1}\Pi_{ij}{\partial \over \partial X_j}\langle  C \rangle
~~~.
\end{equation}
where
\begin{equation}\label{12}
\Pi_{ij}    =\int d t^\prime \int d\vec{x}^\prime \langle  G\delta
V_i(\vec{x}, t) \delta V_j(\vec{x}^\prime, t^\prime) \rangle~~~,
\end{equation}
and the Green's function obeys an equation
\begin{equation}\label{13}
\left[{\partial \over \partial t}+ \left(\langle \vec{V} \rangle +\delta
\vec{V}\right){\partial \over \partial \vec{x}} - D_M{\partial^2\over
\partial \vec{x}^2}\right] G = \delta
(\vec{x}-\vec{x}^\prime)\delta(t-t^\prime)~~. \end{equation}
Eq.(9) describes the evolution of $~\langle C \rangle ~$ at the  times of the
order of $\v ^{-1}~$. It illustrates a rather obvious fact that, if the mean
velocity exists in turbulent flow, then the main effect is the transport
of an admixture cloud as a whole with this velocity.
 Eq.(11) describes the evolution of $~\langle C \rangle~$  at the times  of the
order of $\v ^{-2}~$. Thus, turbulent flows are essential at the times of
$\v ^{-2}~$.

In general case the tensor $~\Pi_{ij}~$ can be decomposed as sum of
symmetric $~\Pi_{ij}^S \equiv D_{ij}~$ and antisymmetric
$~\Pi_{ij}^A~$ parts. Then, Eq.(11) can be written as
\begin{equation}\label{14}
{\partial \langle C \rangle  \over \partial T_2} +{\partial \over \partial
X_i}\left[W_i\langle C \rangle -D_{ij}{\partial\langle C \rangle \over
\partial X_j}-D_M~{\partial\langle C \rangle \over \partial X_j}\right]=0
\end{equation}
where $~W_i =\partial \Pi_{ij}^A/\partial X_j~$, and the convective and
the diffusion flows are separated.

 Thus, we have demonstrated that the average transport equation at times of
 the order of $~\v ^{-2}~$ contains two main turbulent effects, namely, the
 turbulent diffusion flow $~-D_{ij}\partial \langle C \rangle /\partial
 X_j~$ and the anomalous convective flow $~\vec{W}\langle C \rangle ~$.

 Turbulent diffusion leads to the extension of an admixture cloud, the
 characteristic size of the  cloud grows as a square root of the slow
 time. The convective flow $~\vec{W}\langle C \rangle ~$ leads to the
 transport of the cloud as whole, the deviation of the coordinate of the
 cloud center grows proportionally to time. This circumstance leads to the
 fact that at sufficiently large times the  effect of the anomalous
 convective flow will prevail over the effect of the turbulent diffusion.
 In other words, the characteristic size of the diffusive extension
 becomes smaller than the  deviation of the cloud center due to the
 anomalous convective flow.

\section{Generalized helicity and anomalous convective flow}

 In this Section we show that the convective flow  is non--zero for
 locally homogeneous isotropic reflectionally noninvariant turbulence.

 Usually in a kinematic statement of problems concerning passive admixture
 transport one does not use an explicit form of velocity correlation
 tensor [18]. It is possible due to simple representation of the tensors,
 which are invariant to certain space transformations [3,4] (we mention
 alpha-effect in magnetic dynamo theory as a classic example).
 Let us start with the case of homogeneous isotropic reflectionally
 noninvariant turbulence $~\delta\vec{V}(\vec{x},t)~$. Due to the
 invariance of the statistical characteristics of the turbulence under
 dilatations and rotations, the correlation tensor $~\langle \delta
 V_i\delta V_j \rangle ~$ in general case is a linear combination of
 elementary tensors $~\delta_{ij},r_ir_j(\vec{r}=\vec{x}-\vec{x}^\prime
 )~$ and a pseudotensor $~\v_{ijl}r_l ~$ with the coefficients depending
 on $~|\vec{r}|~$ (it is also invariant under transformations mentioned
 above). It is important to note that the coefficients at $~\delta_{ij}~$
 and $~r_ir_j~$ are scalars while the coefficient at $~\v_{ijl}r_l~$ is a
 pseudoscalar. This fact is a consequence of the invariance of $~
 \langle \delta  V_i\delta V_j \rangle  ~$ also under the reflections of
 coordinate axes. Nevertheless, in spite of the fact that
 $~\langle \delta  V_i\delta V_j \rangle ~$ is invariant under the
 reflections, the pseudoscalar in it is also the correlation moment of the
 velocity field. In this case it coincides with the helicity. This moment
 is noninvariant under the reflections. The turbulence with nonzero
 correlation moments noninvariant under the axes reflection can be
 naturally called reflectionally noninvariant. As a matter of fact, this
 property extracting the pseudoscalar coefficient can be regarded as
 another definition of the helicity.

 Further on we shall be interested in the explicit form of the tensor
 $~\Pi_{ij}~$. Let us consider the  tensor
 $~\langle G \delta  V_i \delta  V_j \rangle ~$
 in $~\Pi_{ij}~$. The stochastic field $~\delta
 \vec{V}~$ is homogeneous and isotropic as before. However, the Green's
 function $~G~$ depends on the vector $~ \langle \vec{V} \rangle ~$,
 therefore, the general form of the tensor $~\langle G \delta  V_i \delta  V_j \rangle ~$
 is more comples than the tensors discussed previously. This case is
 close to that of axisymmetric turbulence [24]. In other words, the number
 of elementary tensors which determine the general form of
 $~\langle G \delta  V_i \delta  V_j \rangle ~$ grows.Writing all possible
 elementary tensors and pseudotensors, namely, $~
 \delta_{ij},r_ir_j,r_l\langle V_j\rangle~ ,
 \langle V_i\rangle\langle V_j\rangle, \v _{ijl}r_l,\v _{ijl}
 \langle V_l\rangle~$ etc., see [24], with coefficients depending on $~
 |\vec{r}|, \left |\langle \vec{V}\rangle \right  |~$ and $~ \vec{r}\cdot
 \langle \vec{V}\rangle~$(invariant combinations of the vectors $~\vec{r}~$
 and $~\langle \vec{V}\rangle~$) we get a general form of the tensor
 $~\langle G \delta  V_i \delta  V_j \rangle ~$. After integration over
 space coordinates we get an expression for the tensor $~\Pi_{ij}~$ :
 \begin{equation}\label{15}
 \Pi_{ij} = \int_0 ^\infty  d\tau \left\{A_i\delta_{ij}+A_2
 \langle V_i \rangle  \langle V_j \rangle +H\v _{ijl}\langle V_l \rangle
 \right\}~~,
 \end{equation}
where $~A_1,A_2~$ are scalar functions of $~\left|\langle \vec{V}\rangle
\right|~, \tau~$, while $~H~$ is a pseudoscalar function. The terms with
$~A_1,A_2~$ are symmetric on indices $~i, j~$, and they determine the
turbulent diffusion, that is, the term $~D_{ij}~$ in Eq.(14). If the mean
velocity is equal to zero, then only the term with $~A_1~$ retains in
Eq.(15). If non--zero mean velocity exists, then the term with $~A_2~$
appears in the turbulent diffusion coefficient. This term describes the
influence of the mean velocity on the diffusion processes. However, the
last term in Eq.(15) leads to more radical consequences. This term is
antisymmetric on indices $~i, j~$ and, as it follows from the
consideration above, leads to an anomalous convective flow:
\begin{equation}\label{16}
W_1 = {\partial D_{ij}^A \over \partial X_j} =
G_1 \left(
\nabla \left|\langle \vec{V}\rangle \right|\times \langle \vec{V}\rangle
 \right)_i +G_2 curl_i\langle \vec{V}\rangle ~~,
\end{equation}
where
\begin{equation}\label{17}
G_1 \left(\left|\langle \vec{V}\rangle \right|\right) = \int_0^\infty
d\tau {\partial \over \partial \left|\langle \vec{V}\rangle \right| }
H\left( \left|\langle \vec{V}\rangle \right| , \tau \right)~~~,
\end{equation}

\begin{equation}\label{18}
G_2 (|\langle \vec{V}\rangle |) = \int_0^\infty
d\tau
H (|\langle \vec{V}\rangle | ,\tau )~~~.
\end{equation}
\vspace{3mm}

Let us turn to the physical meaning of the pseudoscalar $~H~$. The nonzero
value of $~H~$ implies the existence of some correlation
moments of the turbulent field $~\delta \vec{V}~$  such that these moments
change sign under the reflection of coordinate axes. Just these moments
form the quantity $~H~$.
In this sense $~H~$ is a measure of reflection noninvariance of the
turbulence considered.  Therefore, we call it generalized helicity because
the "ordinary" helicity possesses the same property. Moreover, if we
expand the Green's function in the orders of $~\delta \vec{V}~$, then in
the zero-order approximation $~H~$ is proportional to the ordinary
helicity density.

Thus, the effect of anomalous convective transport appears in
reflectionally noninvariant turbulence, and the value of convective
transport is determined by the  generalized helicity which, in part, is
non--zero for the turbulence possessing an ordinary helicity.

An explicit expressions for the turbulent diffusion and the anomalous
convective flow can be obtained in various limit cases by solving the
equation (13) for the Green's function by different methods, e.g., the
Green's function of zeroth order approximation on fluctuations  serving
as bare one; with using the assumption of the Gaussian character of the
velocity  fluctuations and the Furutzu--Novikov formula; with the help of
an expansion in small Pecle numbers, etc. The simplest expressions can be
obtained by neglecting velocity fluctuations and molecular diffusion in
Eq.(13) (it corresponds to the zeroth order approximation of the diagram
technique). Then the Green's function takes the form
\begin{equation}\label{19}
G = \theta(t-t^\prime) \delta (x-x^\prime - \langle\vec{V}\rangle
(t-t^\prime))
\end{equation}
and $\Pi_{ij} $
\begin{equation}\label{20}
\Pi_{ij} = \int^t_0 \d\tau \langle\delta V_i \delta V_j \rangle
 _{\langle\vec{V}\rangle \tau,\tau} \end{equation}

Eq.(11) with $\Pi_{ij}$    given by Eq.(20) has been obtained previously 
[26] with using assumption that the turbulent velocity is a telegraph 
stochastic process. Eqs.(19), (20) describe turbulent diffusion in 
homogeneous isotropic nad anisotropic turbulence as well.

Inserting two-point correlation tensor of homogeneous isotropic turbulence
[3,4] into Eq.(20) one can easily see that the generalized helicity is
reduced to an ordinary one, whereas in Eq.(16)-(18)

$$ H\left( \left|\langle \vec{V}\rangle \right| , \tau \right) \to
g \left( \left|\langle \vec{V}\rangle \right| \tau , \tau \right) $$
where
$$ g(0,0) = {1\over 6} \langle \delta \vec{V}(\vec{x},t)\cdot curl
\delta \vec{V}(\vec{x},t)\rangle $$
coincides (up to numerical factor) with the "ordinary" helicity density.

The effect considered is the most interesting in the cases when the
direction of the anomalous convective flow is orthogonal to the direction
of the flow $\langle \vec{V}\rangle \langle C \rangle $ :
\begin{enumerate}
\item The velocity gradient is perpendicular to the velocity direction.
Such a situation is rather frequent; the boundary layer of the atmosphere
can be mentioned, when the wind velocity is parallel to the surface while
the gradient of the velocity is perpendicular to the surface. Then, the
anomalous convective flow is directed "sideways", that is, along the third
coordinate parallel to the surface. It is interesting to note the analogy
of this effect and the effect previously described by us [25] : a particle
beam with an inhomogeneous velocity profile in an external stochastic
helical field generates a new particle flow in the transverse direction.

Therefore, in a homogeneous isotropic reflectionally noninvariant
turbulence with the gradient of the mean velocity the drift of an
admixture in the direction perpendicular to the wind direction arises
along with the diffusive extension.

\item The mean flow $\langle \vec{V}\rangle $ is a large--scale plane
vortex. Such vortices appear, for example, when a wind flows around an
obstacle. In this case the anomalous convective flow is directed away
from the surface or towards it in accordance with the sign of the helicity
density. Therefore, in a laminar plane vortex flow the mechanism of
lifting or settling of an admixture exists due to the reflection
noninvariance of the homogeneous isotropic turbulence.

The consideration presented above  demonstrates an importance
of taking into account the property of reflection noninvariance of the
turbulence when studying the effects of turbulent transport. Astrophysical
and geophysical phenomena with a very large range of scales can provide
the existence of the effects arising due to anomalous convective
processes.
\end{enumerate}

\section{Results }

In the paper we study the transport of a passive admixture in a helical
(reflectionally noninvariant) turbulence with a weakly inhomogeneous mean
flow.\\
\begin{enumerate}

\item  With the use of multiscale formalism a closed equation is obtained
which describes the transport of passive admixture in such a turbulence.
\item  The dependence of the turbulent diffusivity tensor on the mean flow
velocity is obtained.
\item  It is shown that a new effect arises in this turbulence, namely,
the anomalous convective transport. This effect comes in the same order
as the turbulent diffusion.
\item  It is demonstrated that the direction of the anomalous convective
flow does not coincide with the direction of the mean flow. Natural
physical cases are presented when the direction of the anomalous flow is
perpendicular to the direction of the mean flow.
\item  It is proved that the violence of reflection invariance of the
turbulence possessing a weakly inhomogeneous flow manifests itself just in
the appearance of the anomalous convective transport.
\end{enumerate}
\vspace{5mm}

{\bf Acknowledgements }\\

This paper was supported by International Association under the project
INTAS--93--1194 and by the State Committee on Science and Technologies of
Ukraine, grant No 2/278.


\newpage

\begin{thebibliography}{99}

\bibitem{1} A.S.Monin, A.M.Yaglom, Statistical Fluiid Meshanism. MIT Press
(Cambridge, Mass.), 1971 (1965 in Russion).
\bibitem{2} D.C.Leslie,
Developments in the Theory of Turbulence. Clarendon Press (Oxford), 1973.
\bibitem{3}G.K.Batchelor, The Theory of Homogeneous Turbulence
(Cambridge), 1953.
\bibitem{4}L.F.Richardson , Proc.Royal.Soc.London, {\bf
A 110} (1926) 709.
\bibitem{5}A.N.Kolmogorov, Cr.Akad.Sci. (USSR) {\bf
30} (1941) 301.
\bibitem{6}B.B.Mandelbrot, The Fractal Geometry of Nature, W.N.Freeman
(NY), 1982.
\bibitem{7}U.Frisch, P.L.Sulem, M.Nelkin, Journ.Fluid Mech., {\bf 87}
(1978) 719.
\bibitem{8}J.L.McCauley, Phys.Reports, {\bf 198} (1990) 226.
\bibitem{9}S.Grossman, I.Procaccia, Phys.Rev. {\bf A 29} (1984) 1358.
\bibitem{10} H.G.E.Hentschel, I.Procaccia, Phys.Rev. {\bf A 29}
(1984)1461.
\bibitem{11} V.Yakhot, S.A.Orszag, Journ.Sci.Comput. {\bf 1}
(1986) 3.
\bibitem{12} W.P.Dannevik, V.Yakhot, S.A.Orszag, Phys.Fluids,
{\bf 30} (1987) 1.
\bibitem{13} R.H.Kraichnan, Phys.Fluids, {\bf 13}
(1970) 22.
\bibitem{14} Ya.B.Zeldovich, Dokl.Akad.Nauk SSSR {\bf 226}
(1982) 821; (Sov.Phys.Dokl. {\bf 27} (1982) 797).
\bibitem{15} U.Frisch,
Lectures on Turbulence and Lattice Gas Hydrodynamics. In: Lecture Notes on
Turbulence. NCAR--GTP Summer School, June 1987, p. 219.
\bibitem{16} S.S.Moiceev, R.Z.Sagdeev, A.V.Tur, G.A.Khomenko, V.V.Yanovsky,
Zhurn. Eksper.Teor.Fiz. (Sov.JETP) {\bf 58} (1983) 1144.
\bibitem{17} J.J.Moreau, C.R. Akad.Sci.Paris, {\bf 252} (1961) 2810.
\bibitem{18} H.K.Moffat, Magnetic~ Field~ Generation in Electrically
Conducting~ Fluids,\\ Cambridge Univ.Press. (Cambridge), 1978.
\bibitem{19} A.V Tur, V.V.Yanovsky, Journ.Fluid.Mech. {\bf 248} (1993) 67.
\bibitem{20} S.S.Moiseev, P.B.Rutkevich, A.V Tur, V.V.Yanovsky,
Zhurn. Eksper. Teor. Fiz.  (Sov.JETP) {\bf 94} (1988) 144.
\bibitem{21} R.Z.Sagdeev,  S.S.Moiseev, P.B.Rutkevich, A.V Tur,
V.V.Yanovsky,~~  In:  Tropical Meteorology. Proc. 3th Int.Symp. Leningrad,
Gidrometeoizdat, 1987, p. 18.  \bibitem{22} V.G.Levich, Hydrodynamics of
Chemical Physics (in Russian).  M., Acad. Sci. USSR Publ., 1952.
\bibitem{23} Al.H.Nayfeh, Perturbation Methods, Y.Wiley \& Sons, Inc.,
1973.  \bibitem{24} G.S.Chandrasekhar, Phil.Trans.Eoy.Soc. {\bf A 242} No
855 (1950) 557.  \bibitem{25} A.V.Chechkin, A.V Tur, V.V.Yanovsky, Physica
{\bf A 208} (1994) 501.

\end{thebibliography}
\end{document}